\documentclass[sn-mathphys-num,iicol]{sn-jnl}

\usepackage{graphicx}%
\usepackage{multirow}%
\usepackage{amsmath,amssymb,amsfonts}%
\usepackage{amsthm}%
\usepackage{mathrsfs}%
\usepackage[title]{appendix}%
\usepackage{xcolor}%
\usepackage{textcomp}%
\usepackage{manyfoot}%
\usepackage{booktabs}%
\usepackage{algorithm}%
\usepackage{algorithmicx}%
\usepackage{algpseudocode}%
\usepackage{listings}%

\theoremstyle{thmstyleone}%
\theoremstyle{thmstyletwo}%

\theoremstyle{thmstylethree}%

\def\be{\begin{equation}}
	\def\ee{\end{equation}}

\newcommand{\beq}{\begin{eqnarray}}
	\newcommand{\eeq}{\end{eqnarray}}
\def\be{\begin{equation}}
	\def\ee{\end{equation}}
\newcommand\eq[1]{Eq.~(\ref{#1})}

\begin{document}

\title[Article Title]{Casimir versus Helmholtz fluctuation induced force in { the Nagle}-Kardar model: Exact results }


\author*[1,2]{\fnm{Daniel} \sur{Dantchev}}\email{danieldantchev@gmail.com}

\author[2]{\fnm{Nicholay} \sur{S. Tonchev}}\email{nicholaystoichev@gmail.com}
\equalcont{These authors contributed equally to this work.}

\author[3]{\fnm{Joseph} \sur{Rudnick}}\email{jarudnick@physics.ucla.edu}
\equalcont{These authors contributed equally to this work.}

\affil*[1]{\orgdiv{Institute of Mechanics}, \orgname{Bulgarian Academy of Sciences}, \orgaddress{\street{Akad. G. Bonchev St. bl. 4}, \city{Sofia}, \postcode{1113}, \country{Bulgaria}}}

\affil[2]{\orgdiv{Institute of Solid State Physics}, \orgname{Bulgarian Academy of Sciences}, \orgaddress{\street{72, Tzarigradsko Chaussee, Blvd.}, \city{Sofia}, \postcode{1784},  \country{Bulgaria}}}

\affil[3]{\orgdiv{Department of Physics and Astronomy}, \orgname{University of California}, \orgaddress{ \city{Los Angeles}, \postcode{90095}, \state{California}, \country{USA}}}


\abstract{When used to describe \textit{finite} systems the  {conjugate} statistical-mechanical ensembles are \textit{not} equivalent. This has physical implications for the behavior of the fluctuation induced forces pertinent to the different ensembles. Here, {we study the Nagle-Kardar model within the grand-canonical ensemble (GCE) and the canonical ensemble (CE) (with conserved total magnetization) for periodic boundary conditions (PBC)}. {We focus on two fluctuation-induce forces: the Casimir force (CF) in the GCE and the Helmholtz force (HF) in the CE}.   In the infinite system limit the  model exhibits a critical line, which ends at a tricritical point.  Unexpectedly,  the critical Casimir force (CCF) is \textit{repulsive} near the critical line and  tricritical point,   decaying rapidly upon departure from those two regions and becoming \textit{attractive}.  This violates  the widely-accepted ``boundary condition rule,'' which presumes  that the CCF is attractive for equivalent boundary conditions (BC)  and repulsive for conflicting BC. For the HF we find that it also changes sign as a function of temperature and the magnetization. We conclude, that CCF and HF have a behavior quite different from each other as a function of the tunable parameters (temperature, magnetic field, or magnetization) of the model. This dependence allows {for the control of the}   \textit{sign} of these forces, as well as their magnitude. 


	
}

\keywords{fluctuation induced forces, Casimir force, Helmholtz force, phase transitions and critical phenomena}



\maketitle


\section{Introduction}\label{sec:Introduction}

In {recent} years the fluctuation induced forces (FIF) {have been the} object of intense studies{; consequently, there is now}  a vast amount of literature devoted to them. One of the topics intensively studied {has been} FIF in statistical-mechanical systems undergoing continuous phase transitions. Few of the recent reviews on the this subject are \cite{MD2018,DD2022,Gambassi2023,Dantchev2024b}.

The FIF are pertinent  to confined systems {for which boundary conditions at the limiting surfaces play an important role. {Conversely,} taking BC into account  usually greatly complicates the mathematical problem,  even {in the case as such simple cases as the one-dimensional Ising model.} \cite{Ising25,DR2022,Dantchev2023b,Dantchev2024a,Dantchev2024}  and {the} Nagle - Kardar (NK) model \cite{nagle1970ising,kardar1983crossover,DTR2024-PRE,Dantchev2024d}, { the object of the present study. }

To give a flavor of the problems {noted above}, we {we note} two preliminary comments concerning \textit{equivalence } of ensembles and specifics of \textit{long-range interactions}. 

When the influence of finite-size effects are negligible, the standard method in statistical  mechanics is to replace {a} large, but finite system, by {an} infinitely large  one, by  setting $N, V \to \infty$, where $N$ is the number of particles, $V$ is its volume, and the limit is taken so that $N/V = \rho ={\cal O}(1)$,  where $\rho$ is 
the density. Mathematically, this procedure is called ``finding  the thermodynamic limit" \cite{K2007,PB2011, minlos2000,Dorlas2021}. On the other {hand}, when the correlation length is comparable to the size of {a} system, finite-size effects are essential and can not be neglected. In this situation the properties of the system are described by  finite-size scaling theory, see, e.g., \cite{BDT2000,P90}. 

{Directed } by the experimental situation, one considers a {suitable} ensemble for a  system, and  then  calculates the thermodynamic quantities by taking average over this ensemble. Different ensembles  are usually equivalent in the thermodynamic limit, i.e.,  the predictions of statistical mechanics do not depend on the chosen ensemble. There is a practical usefulness  of ensemble equivalence since, for computational purposes, one has the freedom to choose the ensemble in which the calculations are the easiest. 
It is very important to stress that the equivalence statement is true
only for short-range interactions, i.e.,  excluding systems involving
long-range non-additive interactions., see, e.g., \cite{Mori2011,Mori2013}.   An example of a model with such interactions is the NK model, in which, as we will see below, there are long-range equivalent-neighbor ferromagnetic interactions of strength $J_{l}/N>0$
between all the spins. For this model there is {an} abundance of literature, see e.g. \cite{mukamel2005breaking,campa2009statistical,CDR2009,campa2014physics,gupta2017world} and references therein. As it was shown in the canonical ensemble with fixed magnetization, one observes  non-convex behavior of the Helmholtz free energy which  {allows for} stable, metastable and unstable states in the system \cite{Dantchev2024d}.

Ensemble equivalence is  based on certain asymptotic  properties of the partition functions. Equivalence is then a consequence of the concavity of the corresponding thermodynamic potentials {green}{that are} related by the Legendre-Fenchel transform (see e.g. \cite{touchette2005legendre}). 
                                                                                                                                                                                 Lack of equivalence means that this transformation is not involutive, i.e., for example if applied to the bulk Gibbs free energy $f(h,t)$ the result {does} not coincide with the bulk Helmholtz free energy $a(m,t)$, but to {the} convex envelope of $a(m,t)$. Despite that the last is a well known textbook result, we note that in the context of  the Casimir and Helmholtz forces it causes additional (in conjunction  with the presence of finite-size boundary conditions) challenging complications. 

The work reported in the current paper scrutinizes and extends the  observations of  the authors
on the FIF in the NK model  reported in Refs. \cite{DTR2024-PRE,Dantchev2024d}. We start by first recalling the definitions of the CCF and the CF. 

\section{On the definitions of the CCF and HF}
We envisage a  $d$-dimensional system with a film geometry $\infty^{d-1}\times L$, $L\equiv L_\perp$, and with boundary conditions $\zeta$ imposed along the spatial direction of finite extent $L$.  Take ${\cal F}_{ {\rm tot}}^{(\zeta)}(L,T,h)$ to be the total free energy of such a system within the GCE, where $T$ is the temperature and $h$ is the magnetic field. Then, if   $f^{(\zeta)}(T,h,L)\equiv \lim_{A\to\infty}{\cal F}_{ {\rm tot}}^{(\zeta)}/A$  is the free energy per area $A$ of the system, one can define the Casimir force for critical systems in the grand-canonical $(T-h)$-ensemble, see, e.g. Ref. \cite{Krech1994,BDT2000,MD2018,DD2022,Gambassi2023}, as: 
\begin{equation}
	\label{CasDef}
	\beta F_{\rm Cas}^{(\zeta)}(L,T,h)\equiv- \frac{\partial}{\partial L}f_{\rm ex}^{(\zeta)}(L,T,h)
\end{equation}
where
\begin{equation}
	\label{excess_free_energy_definition}
	f_{\rm ex}^{(\zeta)}(L,T,h) \equiv f^{(\zeta)}(L,T,h)-L f_b(T,h)
\end{equation}
is the so-called excess free energy, per area and per $\beta^{-1}=k_B T$, over the bulk free energy density $f_b$.

Along these lines, we define the corresponding  fluctuation induced   Helmholtz force \cite{DR2022,Dantchev2023b} in the canonical $(T-M)$-ensemble, where $M$ is the fixed value of the total magnetization:
\begin{equation}
	\label{HelmDef}
	\beta F_{\rm H}^{(\zeta)}(L,T,M)\equiv- \frac{\partial}{\partial L}f_{\rm ex}^{(\zeta)}(L,T,M)
\end{equation}
and
\begin{equation}
	\label{excess_free_energy_definition_M}
	f_{\rm ex}^{(\zeta)}(L,T,M) \equiv f^{(\zeta)}(L,T,M)-L f_H(T,m).
\end{equation}
Here, the average magnetization $m=\lim_{L, A\to \infty}M/(LA)$, and $f_H(T,m)$ is the Helmholtz free energy density of the ``bulk'' system. 

In the remainder of this article we will take $L=N a$, where $N$ is an integer number, and for simplicity we 
set $a=1$, i.e., all lengths will be measured in units of the lattice spacing $a$. 

Now we pass to the results for the NK model. 

\section{The Nagle-Kardar model}

\begin{itemize}
	
	\item{\textit{The model in the grand canonical ensemble (GCE)}}

For convenience of the reader we recall that the NK model is defined by the Hamiltonian 
\begin{equation}
	\label{NKGC}
	\beta{\cal H}^{{\rm GCE}}=-K_s\sum_{\langle i,j \rangle}^{N}S_iS_j + h\sum_{i=1}^{N}S_i-\frac{K_l}{N}\sum_{i,j=1}^{N}S_iS_j,
\end{equation}
 where $h\in \mathbb{R}$, $\beta=1/(k_B T)$,
\begin{equation}
\label{eq:def-interaction-constants}
K_s=\beta J_s \in \mathbb{R}, \quad K_l=\beta J_l \in \mathbb{R}^+,
\end{equation}
$\langle i,j \rangle$ stands for ``nearest - neighbors'', and we suppose  periodic boundary conditions. 
The first terms are as in the standard Ising model with spins $S_i=\pm 1, i=1,\cdots,N$, with  $N$ being the number of spins, and interaction constant $J_s$, while the last term is the well-known``equivalent neighbors'' long-ranged interaction Husimi-Temperley model characterized by $J_l$.  In the inverse temperature $\beta=1/T$ we will set the Boltzmann constant
$k_B=1$. 

\item{\textit{The model in the canonical ensemble (CE)}}
\begin{eqnarray}
	\label{NKC}
	\beta{\cal H}^{{\rm CE}}&=&\left[-K_s\sum_{\langle i,j \rangle}^{N}S_iS_j   -\frac{K_l}{N}\sum_{i,j=1}^{N}S_iS_j \right] \nonumber \\
	&& \times \; \delta\left(\sum_{i=1}^{N}S_i -M\right),\; \quad M \in \mathbb{R}.
\end{eqnarray} 
Given the symmetry of the problem it suffices to fix magnetization $M\geq 0$.

\end{itemize}

Here it is necessary to make some clarification of the terminology used in the literature. Here  there exist a controversy in the terminology. In  the  literature  devoted to lattice spin models (like Ising) the notation canonical ensemble is often  used
when  the sum in the partition function is taken over all possibles microscopic states without any constraint (see, e.g K. Huang \cite[Ch. 14]{KH2001} and one terms the corresponding free energy the Helmholtz one. However, in other articles, e.g., in E. Stanley \cite{stanley1971phase} one speaks about Gibbs free energy. In the context of the NK model the notion "canonical" in the above sense also uses, see e.g. \cite{CDR2009,bouchet2010thermodynamics,campa2014physics,gupta2017world}.
In the present study we refer to the lattice gas language and perceive the grand canonical ensemble as the one where the magnetization is not
fixed and is
associated to the Gibbs magnetic free energy (see \eq{NKGC}). The notion  "canonical ensemble" is attributed to a partition function 
given by the sum of states  where a constraint is enforced by employing the Kronecker
delta function, say, for the total magnetization, e.g., via $\delta[\sum S_i=M=N m]$, where the $m \in [-1,1]$ (see \eq{NKC}). Connection with the thermodynamics  in the last case is then associate to the Helmholtz
magnetic free energy. For an useful   discussion on the  theme with quotation literature one can  consult Refs. \cite{BLH95,WAP2017,plascak2018ensemble}. 

\subsection{On the phase diagram of the model in the two ensembles}

The phase diagram of the model in GCE has been determined in Ref. \cite{DTR2024-PRE}, while the one for the CE in Ref. \cite{Dantchev2024d}. The combined plot showing both of them is shown in Fig. \ref{fig:phase_diagram}. The full red line there $T/J_l=2[x/W_p(x)] $, with $W_p$ being the principal branch of the Lambert $W$-function \cite{NIST2010}, represents a line of critical points, existing in the both ensembles. At this
line, the spontaneous magnetization critical exponent $\beta=1/2$ \cite{nagle1970ising}.  The full blue line represents a line of first-order phase transition pertinent to the GCE, which ends at $x=-0.5$, while the dotted blue line is such a line in CE. This dotted line ends at $x=0$. At these two lines three phases with the
same free energy, and magnetization $m = 0, m =m_{\pm} \ne 0$, coexist. The green point marks the tricritical point  with coordinates  $\{y_{\rm TP}=2/\sqrt{3}, x_{\rm TP}=-\ln(3)/(2\sqrt{3})\}$. At it $\beta=1/4$. The orange point shows where the tangent to the dotted curve becomes along the vertical axis.  

The fact that the phase diagrams in both ensembles are \textit{not} the same demonstrates that the two ensembles are \textit{not} equivalent, even in the thermodynamic limit. In the GCE all states corresponding to  fixed pair $\{K_s,K_l\}$   are stable as a function of $h$, with positive susceptibility and convex Gibbs free energy. In the CE this is \textit{not} always the case. This question has been studied in some details in \cite{Dantchev2024d}. The Helmholtz free energy, as shown there, is not always concave and there are stable, metastable and unstable states in the system --- as a function of $m$ for given  $\{K_s,K_l\}$. Numerically we observe that the metastable states are bounded by the blue dotted line and the red one, while below the dotted line they are unstable. 
\begin{figure}[htbp]
	\begin{center}
		\includegraphics[width=\columnwidth]{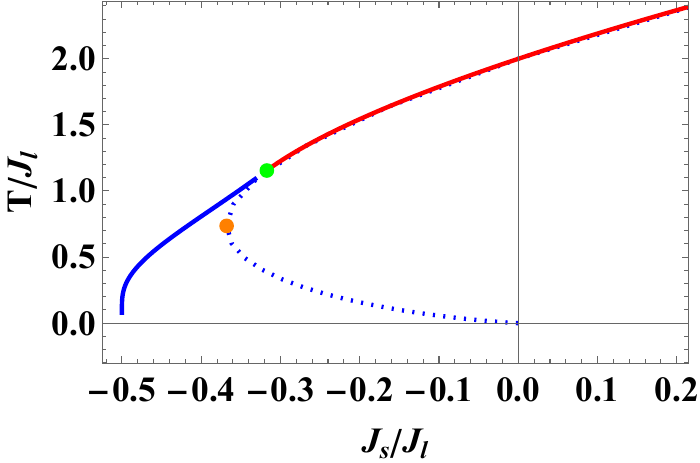}
		\caption{The phase diagram in terms of the renormalized temperature is shown as a function of the ratio $x=J_s/J_l$. The green dot indicates the location of the tricritical point, and the dotted curve delineates a region in which metastability is allowed in the context of the canonical ensemble and is forbidden in the grand canonical ensemble.
			}
		\label{fig:phase_diagram}
	\end{center}
\end{figure}

\subsection{On the behavior of the FIF in the two ensembles}

In \cite{DTR2024-PRE}  it is shown that, in terms of $K_s$ and $K_l$, the critical line is determined by the simple equation $2K_l=\exp(-2K_s)$. The behavior of Gibbs free energy in the finite NK chain using the (generalized) Laplace method, as explained in \cite{DTR2024-PRE}, {yields} 
\begin{eqnarray}
	\label{eq:Z-per-bc-N-large-at-critical-line}
	&& Z_N^{(\rm GC)}(K_s,K_l=e^{-2 K_s}/2,h=0) \\
	&=& \frac{1}{\sqrt{4\pi K_l}} \frac{\Gamma \left(\frac{1}{4}\right) N^{1/4}}{\sqrt[4]{2 e^{2 K_s} \left(e^{4K_s}-1/3\right)}} \nonumber
	\\ 
	&& \times \exp\left\{ N \left[K_s+\ln \left(1+e^{-2 K_s}\right)\right] \right\} \nonumber \\
	&& \times \Bigg \{1+\frac{\sqrt{2\pi K_l}\sqrt[4]{2 e^{2 K_s} \left(e^{4K_s}-1/3\right)}}{\Gamma \left(\frac{1}{4}\right)N^{1/4}}  \nonumber\\
	&& \exp\left\{ -2 N \tanh^{-1}\left(e^{-2 K_s}\right)\right\}\left[1+{\cal O}\left(N^{-1/4}\right)\right] \Bigg \}. \nonumber
\end{eqnarray}
From {this}  follows Eq. (18) {as} reported in \cite{DTR2024-PRE}, which leads to the Casimir amplitude at the critical line $\Delta_{\rm Cas}^{(\rm cr)}=1/4$. Analogically, at the tricritical point $\{K_s=-\ln 3/4,K_l=\sqrt{3}/2,h=0\}$ one has
\begin{eqnarray}
	\label{eq:Z-per-bc-N-large-at-the-tricritical-point}
	&& Z_N^{(\rm GC)}(K_s=-\ln 3/4,K_l=\sqrt{3}/2,h=0)= \nonumber\\
	&&\frac{\sqrt[6]{5/2}}{\sqrt{\pi} 3^{5/6}} \Gamma \left(\frac{1}{6}\right) N^{1/3} \nonumber
	\exp{\left\{N \left[\ln \left(\sqrt{3}+1\right)-\frac{\ln (3)}{4}\right]\right\}}\\ && \Bigg \{1+\frac{(-1)^N}{ N^{1/3} }\frac{\sqrt{\pi} 3^{5/6}}{\sqrt[6]{20} \Gamma \left(\frac{1}{6}\right) }  \\
	&& \times  \exp\left\{ - 2 N \coth ^{-1}\left(\sqrt{3}\right) \right\}\left[1+{\cal O}\left(N^{-1/3}\right)\right] \Bigg \}. \nonumber 
\end{eqnarray}
\begin{figure}
	\begin{center}
		\includegraphics[width=3.0in]{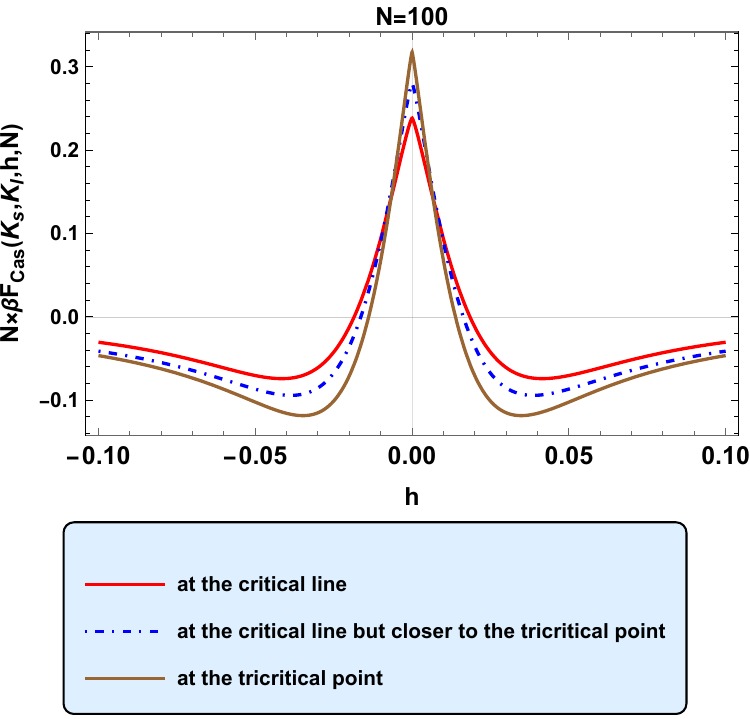} 
		\caption{The behavior of the Casimir force as a function of $h$ for different fixed values of $K_s$ and $K_l$ with $N=100$.  }
		\label{fig:Cas-as-f-of-h}
	\end{center}
\end{figure}
From this expression one derives Eq. (19), reported in \cite{DTR2024-PRE}. From it one also derives the tricritical Casimir amplitude $\Delta^{\rm (tr)}=1/3$. The behavior of the Casimir force as a function of $h$ for three pairs of $\{K_s,K_l\}$ are shown in Fig. \ref{fig:Cas-as-f-of-h}. The red line corresponds to $K_s=0, K_l=0.5$,  the brawn line corresponds to the tricritical point $K_s=-\ln 3/4,K_l=\sqrt{3}/2$, while the blue dash-doted line corresponds
to $K_s=-0.235, K_l=0.800$, i.e., closer to the tricritical point. As we observe, when the point on the critical line is closer to the tricritical one the maximum of the Casimir force increases and approaches the value of $\Delta^{\rm (tr)}$. 

The behavior of the critical Casimir force at the critical line as a function of $K_l\in (0,2]$ for $N=100, 200, 400$ and $N=800$, is shown in Fig. \ref{fig:on-the-critical-line}. Particularly, the red curve, for $N=100$, is the section of  the surface depicted in Fig. 3 of Ref. \cite{DTR2024-PRE} with the plane $h=0$. From the current figure one can see, in addition, that the maximum of the force then is about $K_l=0.93$ and is $\approx 0.35$. The minimum is about $K_l=1.16$ and is $\approx -0.196$. The data for other values of $N$ demonstrate that the minimum deepens with  increase of $N$. The position of this minimum also depends on $N$. This issue, as well as the point of the crossover behavior from critical to the tricritical behavior represents problems that await its future considerations. 

In Ref. \cite{Dantchev2024d} we studied in details the fluctuation induced force for the Nagle-Kardar model in the canonical ensemble when the total value of the magnetization in the system is fixed. We found that the corresponding Helmholtz FIF differs from the Casimir force. It {should} be stressed that because of the very simple dependence of the free energy of the finite system on $K_l$ in this ensemble the HF does not depend on $K_l$ since the bulk and the finite contributions in the excess free energy cancel each other. This leads to the fact that the HF in NK model actually equals {that one of} the HF in the Ising model. The behavior of the force as a function of $K_s$ for $N=100$ and for $m=0.1, 0.4$ and $m=0.8$ is shown in Fig. \ref{fig:helmholtz-force}. The figure demonstrates that the force has {deeper} negative minimum for smaller values of $m$. The force is, however, repulsive for all  {displayed} values of $m$ for large $K_s$, as well as for negative values of $K_s$.

\begin{figure}
	\centering
	\includegraphics[width=\columnwidth]{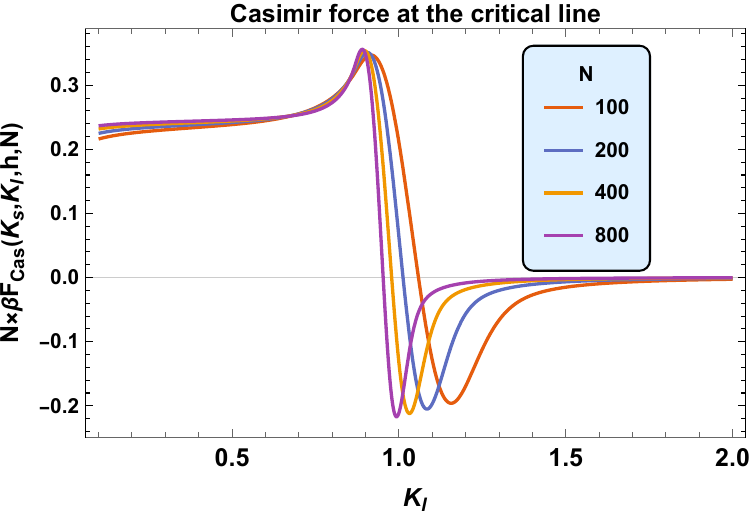}
	\caption{The repulsive and attractive behavior of the critical Casimir force at the critical line as a function of $K_l\in (0,2]$ for $N=100, 200, 400$ and $N=800$. }
	\label{fig:on-the-critical-line}
\end{figure}

\begin{figure}
	\centering
	\includegraphics[width=\columnwidth]{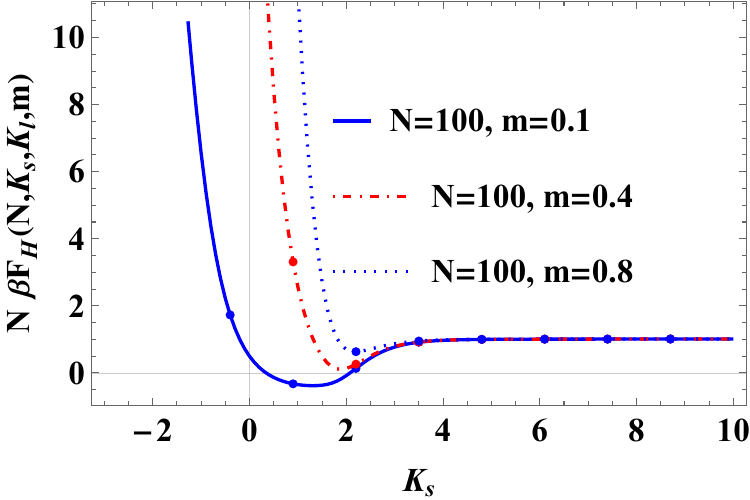}
	\caption{The  behavior of  Helmholtz force $\beta F_{\rm H}(N,K_s,K_l,m)$ as a function of $K_s$ for three fixed values of  $m$ and $N=100$. Let us stress that the force \textit{does not} depend on $K_l$. There is a region where it is repulsive and  increases linearly for negative values of $K_s$. For large positive $K_s$ the three curves show the usual scaling behavior of the one-dimensional Ising model - for more information see \cite{Dantchev2024a,Dantchev2023b,Dantchev2024b}.}
	\label{fig:helmholtz-force}
\end{figure}

\section{Discussion}

For confined systems it has been recently 
shown in the literature \cite{DR2022,Dantchev2023b,Dantchev2024a,Dantchev2024,DTR2024-PRE,Dantchev2024d} that
two ensembles which {impose} different constraints on the fluctuating  parameters lead to qualitatively different behavior of the corresponding FIF. This is due to the {fact} that for finite systems the Gibbs ensembles correspond to quite different physical conditions{; it is then} reasonable 
that the fluctuations
 lead to different FIF, namely  Casimir force (CF) in the grand canonical
ensemble (GCE) and Helmholtz force (HF) in the canonical (CE) one with fixed magnetization \cite{Gambassi2023,Dantchev2024b}.

In the current article we have demonstrated the validity of the above general arguments {green}{in the case} of the NK  model in different ensembles, namely GC and CE (with conserved magnetization). As explained in the introduction, this model has {been} extensively studied in the grand canonical (with fixed external field) and in {the} microcanonical ensemble. It turns out that the model has the same critical line, ending at a tricritical point (we point out that this line can be obtained analytically, as explained in \cite{DTR2024-PRE}), using the Lambert $W$ - function), but differs in its behavior below that point.

In the preceding sections we have seen that the nature of the CF and HF in the NK model (attractive or repulsive) even in the case of symmetric boundary conditions, namely periodic, is less obvious than one {might} expect {from {at} first glance.
It is rather surprising that {the} possibility of {a} sign {change} of the forces has not been appreciated before in the case of other models with symmetric boundary conditions. 
Our findings is not  consistent {with the} lore of  fluctuation{-}induced CF where for
symmetrical boundary conditions an attractive force is found while asymmetrical ones lead to a repulsive force \cite{Gambassi2023,Dantchev2024b,RBM2007,NHC2009}.
This   difference  of the behavior of the Casimir force {can} be attributed {to} the competition between the long-range and short-range forces among the spins in the system.  
This, however, does not hold  in the case of CE since 
the HF in the NK model does \textit{not} depend on the {strength} of {the} long-range interaction  $K_l$. The behavior of this force as a function of $K_s$ for several fixed values of the number of particles in the chain is shown on Fig. \ref{fig:helmholtz-force}. In fact, it equals the Helmholtz force for the one- dimensional Ising model. As it was shown in \cite{Dantchev2023b}
the Helmholtz force in {this}  case also changes sign as a function of the temperature and magnetization and can be attractive or repulsive,
depending on their values.

It is {now} known that long-range interactions cause different anomalous
properties: negative specific heat, long-lived metastable states, and ensemble inequivalence \cite{Mori2011,Mori2013,CDR2009,bouchet2010thermodynamics}. {In general} the ensemble
equivalence is taken for granted for the bulk systems, but the fluctuation induced forces have finite-size origin which enforces different behavior between CF
and HF. 

Certainly,   for fluctuation induced forces it is very instructive to have {at hand}  theoretical studies of canonical or grand canonical ensemble dependence  of systems based on Ginzburg-Landau-Wilson
Hamiltonians. Some results in that direction based on some approximative methods have been obtained
\cite{GVGD2016,GGD2017,RSVG2019}.

We believe that it is difficult to overestimate {the value of }  results about Casimir's and Helmholtz's forces obtained within the framework of exactly solvable models.

\section*{Acknowledgments}
The financial support via Grant No KP-06-H72/5 is gratefully acknowledged.

\section*{Declarations}
Data availability: there are no specific data related to this publications.



\end{document}